\documentclass[aps,prd,reprint,onecolumn,superscriptaddress,nofootinbib,floatfix]{revtex4-2}

\usepackage{amsmath,amssymb,amsfonts,mathtools,bm}
\usepackage{float}
\usepackage{physics}
\usepackage{hyperref}
\usepackage{microtype}
\usepackage{graphicx}
\usepackage{xcolor}
\usepackage{booktabs}
\usepackage{tikz}

\usetikzlibrary{decorations.markings, shapes.geometric, calc,patterns}

\providecommand{\dd}{\mathrm{d}}
\newcommand{\LV}{\mathrm{LV}}

\newcommand{\Mc}{M_{\!\rm kink}}
\newcommand{\calL}{\mathcal{L}}

\begin{document}
	
	\title{One-Loop Quantum Correction to the Kink Mass in a Lorentz-Violating Scalar Field Theory}

	\author{Reza Moazzemi}
	\email{r.moazzemi@qom.ac.ir}
	\affiliation{Department of Physics, Faculty of Science, University of Qom,  Ghadir Blvd., Qom 371614-6611, I.R. Iran}
	
	\begin{abstract}
		We study the radiative correction to the mass of the $(1+1)$-dimensional $\lambda\phi^4$ kink in the presence of an irreducible, spacelike Lorentz-violating interaction. The Lorentz-violating operator is chosen as $-\frac12 c\phi^2(u\!\cdot\!\partial\phi)^2$ with $u^\mu=(0,1)$, which is the lowest-order single-scalar operator considered in this class of models that produces effects that cannot be removed by a linear redefinition of coordinates or the field.  We perform the one-loop renormalization of the underlying quantum field theory. In contrast with Lorentz-invariant $\lambda\phi^4$ theory in two dimensions, the Lorentz-violating theory requires, at this order, both a modified mass counterterm and a nonzero field-strength counterterm. The corrected stability problem contains modified bound and continuum modes, while the continuum phase shift remains unchanged at $O(c)$ with the boundary-condition prescription adopted here. Finally, using the vacuum-subtracted zero-point energy and the counterterm Hamiltonian, we obtain the one-loop kink mass through first order in $c$. We give the finite coefficient in the subtraction prescription implemented in the accompanying thesis calculation, for which the Lorentz-violating contribution is numerically negative.
	\end{abstract}
	
	\maketitle
	
	\section{Introduction}
	Topological solitons provide nonperturbative field configurations whose energies are finite and which interpolate between disconnected sectors of the vacuum manifold. The kink of the real scalar $\phi^4$ model in one spatial dimension is the canonical example and has long served as a laboratory for spontaneous symmetry breaking, collective coordinates, semiclassical quantization, and nonperturbative mass renormalization \cite{Rajaraman1982,MantonSutcliffe2004}. The semiclassical quantization of solitons was developed systematically by Dashen, Hasslacher, and Neveu (DHN) in a series of papers in 1974 \cite{DHN4114,DHN4130}. Their treatment expresses the one-loop energy as the difference between fluctuation spectra in the soliton and vacuum sectors, supplemented by the counterterms required by the underlying quantum field theory.
	
	Lorentz symmetry is a central organizing principle of relativistic quantum field theory. Nevertheless, a violation of particle Lorentz symmetry can be parametrized consistently while preserving observer covariance, most systematically in the framework of Lorentz-violating effective field theories and the Standard-Model Extension \cite{ColladayKostelecky1997,ColladayKostelecky1998,Kostelecky2004}. In scalar theories, both classical and quantum effects of Lorentz violation have been studied, including radiative and Casimir corrections \cite{FerreroAltschul2011,MojaveziMoazzemiZomorrodian2019}.
	
	For a single scalar field in $(1+1)$ dimensions, the lowest-dimensional CPT-even operator $-\frac12 k^{\mu\nu}\partial_\mu\phi\partial_\nu\phi$ can be removed, at leading order, by a linear transformation of coordinates. Consequently it does not generate an intrinsic modification of the single-field kink problem. Barreto, Bazeia, and Menezes emphasized this point in Lorentz- and CPT-violating defect models \cite{BarretoBazeiaMenezes2006}. A non-removable effect appears at the next relevant order when the Lorentz-violating background couples directly to the interacting scalar sector.
	
	The appropriate interaction for the present problem is
	\begin{equation}
		\calL_{\LV}= -\frac12 c\,\phi^2\left(u^\mu\partial_\mu\phi\right)^2,
		\label{eq:LVoperator}
	\end{equation}
	with a constant preferred vector $u^\mu$. Moazzemi, Ettefaghi, and Mojavezi showed that, for the spacelike choice $u^\mu=(0,1)$, this operator changes the classical kink profile and mass and also modifies the bound and continuum fluctuation spectrum \cite{MoazzemiEttefaghiMojavezi2021}. The timelike choice does not alter the static kink.
	
	The remaining conceptual step is quantum consistency. In two-dimensional Lorentz-invariant $\lambda\phi^4$ theory the one-loop renormalization of the vacuum sector is especially simple: at the order relevant for the kink mass, the mass counterterm is the only divergent counterterm. In the present Lorentz-violating theory that statement changes. A one-loop renormalization of the same scalar model shows that a nonzero field-strength counterterm is induced, in addition to a Lorentz-violation-dependent correction to the mass counterterm \cite{ShokriMoazzemi2025}. This difference is physically important because the counterterms contribute directly to the kink-vacuum energy difference.
	
	The purpose of this paper is to combine these ingredients into a single semiclassical calculation of the kink mass through first order in the Lorentz-violating parameter. The structure is deliberately organized so that the renormalization problem is solved before the soliton mass is evaluated. We first define the model and classical kink, then review the ordinary DHN construction, derive the Lorentz-violating counterterms, analyze the corrected fluctuation spectrum, and finally construct the renormalized one-loop mass.\section{Renormalization of the Standard Kink Mass}
	\label{sec:LI}
	
	In this section, we review the renormalization of the kink mass in the canonical $(1+1)$-dimensional $\lambda\phi^4$ scalar field theory. The goal is to compute the one-loop quantum correction to the kink mass using the semiclassical DHN (Dashen–Hasslacher–Neveu) approach, which provides a systematic framework for quantizing topological solitons.
	
	\subsection{Classical Kink Configuration}
	We work in Minkowski spacetime with metric $\eta_{\mu\nu}=\mathrm{diag}(1,-1)$. The Lorentz-invariant theory is
	\begin{equation}
		\mathcal{L} = \frac{1}{2}(\partial_\mu\phi)^2 - \frac{\lambda}{4}\left(\phi^2 - \frac{m^2}{\lambda}\right)^2,
	\end{equation}
	which is invariant under the discrete $Z_2$ symmetry $\phi \to -\phi$. The potential has two degenerate minima at $\phi = \pm m/\sqrt{\lambda}$, which spontaneously break the symmetry. Introducing dimensionless variables
	\begin{equation}
		z = \frac{m x}{\sqrt{2}}, \qquad \varphi = \frac{\sqrt{\lambda}}{m}\,\phi,
	\end{equation}
	the static kink solution, interpolating between the two vacua, is
	\begin{equation}\label{kinkstd}
		\varphi_k^{(0)}(z) = \tanh z.
	\end{equation}
	The classical mass is obtained by integrating the Hamiltonian density,
	\begin{equation}
		M_{\mathrm{cl}} = \frac{2\sqrt{2}}{3}\frac{m^3}{\lambda}.
		\label{eq:Mcl_standard}
	\end{equation}
	
	\subsection{Small Fluctuations and Spectral Problem}
	To quantize the kink, we expand the field about the classical solution:
	\begin{equation}
		\varphi(z,t) = \varphi_k(z) + \eta(z,t),
	\end{equation}
	and retain terms quadratic in $\eta$. The resulting quadratic action yields the stability equation
	\begin{equation}
		\left[ -\partial_z^2 - 1 + 3\tanh^2 z \right] \zeta_n(z) = \omega_n^2 \zeta_n(z).
		\label{eq:stab_standard}
	\end{equation}
	This is a Pöschl–Teller equation whose spectrum is exactly known. There are two bound states:
	\begin{align}
		\text{Zero mode (translational)}: \quad & \omega_0 = 0, \quad \zeta_0(z) = \sech^2 z, \\
		\text{Shape mode}: \quad & \omega_1 = \sqrt{\frac{3}{2}}, \quad \zeta_1(z) = \frac{\sinh z}{\cosh^2 z}.
	\end{align}
	The continuum states are labeled by a continuous parameter $q$:
	\begin{equation}
		\omega_q = \sqrt{2 + \frac{q^2}{2}}, \qquad 
		\zeta_q(z) \sim e^{iqz}\left(3\tanh^2 z - 1 - q^2 - 3iq\tanh z\right),
	\end{equation}
	with the asymptotic phase shift
	\begin{equation}
		\delta(q) = -2\tan^{-1}\left(\frac{3q}{2 - q^2}\right).
	\end{equation}
	Imposing periodic boundary conditions in a box of length $L$ gives the quantization condition
	\begin{equation}
		\frac{q_n L}{\sqrt{2}} + \delta(q_n) = 2\pi n, \qquad n \in \mathbb{Z}.
		\label{eq:q_quant}
	\end{equation}
	
	\subsection{Zero-Point Energy and Divergences}
	The ground-state energy of the kink sector (no excited modes) is
	\begin{equation}
		E_{\mathrm{kink}} = M_{\mathrm{cl}} + \frac{1}{2}\omega_1 + \frac{1}{2}\sum_{n} \omega_{q_n},
	\end{equation}
	where the zero mode ($\omega_0=0$) is excluded. The vacuum sector (constant background $\varphi = \pm 1$) has the energy
	\begin{equation}
		E_{\mathrm{vac}} = \frac{1}{2}\sum_{k} \sqrt{k^2 + 2m^2},
	\end{equation}
	with $k_n = 2\pi n/L$. The physical quantity is the difference:
	\begin{align}
		E_{\mathrm{kink}} - E_{\mathrm{vac}} = &\, M_{\mathrm{cl}} + \frac{1}{2}\omega_1 \nonumber \\
		&+ \frac{1}{2}\sum_{n} \left[ \sqrt{2 + \frac{q_n^2}{2}} - \sqrt{k_n^2 + 2} \right].
		\label{eq:Evac_sub_standard}
	\end{align}
	Using Eq.~\eqref{eq:q_quant} to relate $q_n$ and $k_n$, and taking the continuum limit $L\to\infty$, the sum becomes an integral:
	\begin{equation}
		\frac{1}{2}\sum_{n} \left[ \sqrt{2 + \frac{q_n^2}{2}} - \sqrt{k_n^2 + 2} \right]
		= -\frac{1}{4\pi}\int_{-\infty}^{\infty} dk \, \frac{k\,\delta(k)}{\sqrt{k^2 + 2}}.
	\end{equation}
	After integration by parts, this yields a logarithmic divergence plus a finite part:
	\begin{equation}
		E_{\mathrm{kink}} - E_{\mathrm{vac}} = M_{\mathrm{cl}} + \frac{1}{2}\sqrt{\frac{3}{2}} - \frac{3}{\pi\sqrt{2}} + \text{(log divergence)}.
	\end{equation}
	
	\subsection{Mass Counterterm and the Tadpole Condition}
	In $(1+1)$-dimensional $\phi^4$ theory, super-renormalizability implies that only the mass counterterm is needed at one loop. The mass counterterm is introduced to cancel the divergence in the vacuum two-point function. This is achieved by imposing the \textit{tadpole condition}, which requires that the one-loop tadpole diagram (the contribution to the one-point function) vanishes order by order in perturbation theory. 
	
	In the context of the effective potential, the tadpole condition is expressed as
	\begin{equation}
		\left. \frac{\partial V_{\mathrm{eff}}(\phi)}{\partial \phi} \right|_{\phi = \langle \phi \rangle} = 0,
	\end{equation}
	which ensures that the vacuum expectation value of the field remains at the minimum of the effective potential after including quantum corrections. This condition is equivalent to imposing the renormalization condition on the full propagator, namely that the pole of the propagator is located at the physical mass $m^2$. As discussed in standard textbooks on quantum field theory, such as Peskin and Schroeder \cite{PeskinSchroeder}, the renormalization conditions for the scalar field require that the 1PI two-point function satisfies
	\begin{equation}
		M(p^2)\big|_{p^2 = m^2} = 0,
	\end{equation}
	which fixes the mass counterterm and ensures (and technically equivalent to) that  the tadpole contributions are properly subtracted.
	
	For the present theory, the mass counterterm is defined by
	\begin{equation}
		\delta m = -\frac{3\lambda}{4\pi} \int_{-\infty}^{\infty} \frac{dk}{\sqrt{k^2 + 2m^2}},
	\end{equation}
	which precisely cancels the logarithmic divergence of the tadpole diagram. Its contribution to the kink–vacuum energy difference is
	\begin{equation}
		\Delta E_{\mathrm{ct}} = \frac{1}{2}\delta m \int_{-\infty}^{\infty} dx \left( \varphi_k^2 - 1 \right)
		= \frac{3\sqrt{2}\,m}{4\pi} \int_{-\infty}^{\infty} \frac{dk}{\sqrt{k^2 + 2}}.
		\label{eq:ct_standard}
	\end{equation}
	This expression has the same logarithmic divergence as the zero-point energy difference, but with the opposite sign, so the divergences cancel exactly. The tadpole condition thus guarantees that the renormalized perturbative expansion is well-defined and that the physical mass corresponds to the pole of the propagator.
	
	\subsection{Final Renormalized Mass}
	Adding the counterterm contribution Eq.~\eqref{eq:ct_standard} to the unrenormalized difference Eq.~\eqref{eq:Evac_sub_standard} and taking the finite part yields the renormalized one-loop kink mass:
	\begin{equation}
		M_{\mathrm{kink}} = \frac{2\sqrt{2}}{3}\frac{m^3}{\lambda}
		+ m\left( \frac{1}{2\sqrt{6}} - \frac{3}{\pi\sqrt{2}} \right)
		+ \mathcal{O}(\lambda).
		\label{eq:Mfinal_standard}
	\end{equation}
	This is the well-known DHN result. The one-loop correction is negative and independent of $\lambda$ (except through the classical term), indicating that quantum fluctuations reduce the effective kink mass.
	
	\subsection{
		Remarks on Renormalization in Two Dimensions: Power Counting, Spontaneous Symmetry Breaking, and Counterterms}
	The renormalization analysis is most transparent when it is separated into two logically distinct steps. First, one identifies, by power counting and topology, which 1PI Green functions can contain ultraviolet divergences. Second, one evaluates those divergent functions and determines the local counterterms required to absorb their divergent parts.
	
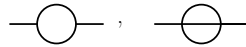
\begin{figure}[H]
		\centering	
		\resizebox{0.18\columnwidth}{!}{%
			\begin{tikzpicture}[line width=1.8pt] % ضخامت خطوط مشابه تصویر اصلی
				% خط ورودی سمت چپ
				\draw (0, 0) -- (1.0, 0);
				
				% دایره مرکزی (حباب)
				\draw (1.7, 0) circle (0.7cm);
				
				% خط خروجی سمت راست
				\draw (2.4, 0) -- (3.4, 0); \node at (4, 0) {\Huge ,};
			\end{tikzpicture}% , 
			\qquad \quad \begin{tikzpicture}[line width=1.8pt] % ضخامت خطوط مشابه تصویر اصلی
				% خط ورودی سمت چپ
				\draw (0, 0) -- (1.0, 0);
				
				% دایره مرکزی (حباب)
				\draw (1.7, 0) circle (0.7cm);
				\draw (1.0, 0) -- (3.4, 0);
				% خط خروجی سمت راست
				\draw (2.4, 0) -- (3.4, 0);
			\end{tikzpicture}
		}
		\caption{ The first two diagrams which contribute in $\delta Z$ in two-point function.}
		\label{fig:rencond}
	\end{figure}

	The key renormalization fact in two dimensions follows already from power counting. For an $n$-point interaction in $d$ spacetime dimensions, a convenient form of the superficial degree of divergence is
	\begin{equation}
		D=d+(d-4)V-\frac{d-2}{2}N,
		\label{eq:Dgeneral}
	\end{equation}
	where $V$ is the number of interaction vertices and $N$ is the number of external lines. For $d=2$ and $n=4$ this becomes
	\begin{equation}
		D=2-2V.
		\label{eq:Dphi4}
	\end{equation}
	Thus the one-loop tadpole, with $D=0$,  is logarithmically divergent, whereas the one-loop four-point function is finite. On the other hand, 	for the broken phase the Lagrangian contains quadratic, cubic, and quartic interactions.The cubic interaction introduces an additional one-loop topology in the two-point function, but its superficial degree of divergence is negative (see please Fig. \ref{fig:rencond}). This diagram, along with the other two-point diagram that contribute to $\delta Z$ at same order, is apparently finite.
	
	Hence the broken-symmetry expansion does not introduce a new ultraviolet counterterm in $(1+1)$ dimensions. In the Lorentz-invariant kink problem the ultraviolet subtraction may equivalently be formulated through normal ordering, in agreement with the standard two-dimensional treatment of the model.
	
	This makes the semiclassical quantization of the kink particularly tractable and the result exact in the sense of the $1/\lambda$ expansion. The absence of a field-strength counterterm $\delta Z$ in the standard theory is a crucial feature that will be contrasted with the Lorentz-violating extension in the following sections.
	
	\section{Irreducible spacelike Lorentz violation}
	\label{sec:LV}
	
	\subsection{LV operator and the classical theory}
	We now add Eq.~\eqref{eq:LVoperator} and choose
	\begin{equation}
		u^\mu=(0,1).
	\end{equation}
	The full Lagrangian is then
	\begin{equation}
		\calL=\frac12(\partial_t\phi)^2-\frac12(\partial_x\phi)^2
		-\frac12c\,\phi^2(\partial_x\phi)^2
		-\frac\lambda4\left(\phi^2-\frac{m^2}{\lambda}\right)^2.
		\label{eq:fullL}
	\end{equation}
	Because $[\phi]=0$ and $[(\partial_x\phi)^2]=2$ in two dimensions, the coefficient $c$ is dimensionless. The operator therefore belongs to the renormalizable effective theory in $(1+1)$ dimensions.
	
	The equation of motion can be written as
	\begin{equation}
		-\partial_x^2\phi+c\phi(\partial_x\phi)^2
		-c\phi^2\partial_x^2\phi-m^2\phi+
		\lambda\phi^3=0.
		\label{eq:EOMLV}
	\end{equation}
	We solve this equation perturbatively,
	\begin{equation}
		\varphi_k(z)=\varphi_k^{(0)}(z)+c\varphi_k^{(1)}(z)+O(c^2).
		\label{eq:kinkexp}
	\end{equation}
	The zeroth-order solution is the standard kink of Eq.~\eqref{kinkstd}. At first order the correction satisfies an inhomogeneous Pöschl--Teller equation and is
	\begin{equation}
		\varphi_k^{(1)}(z)=-\frac12\sech^2z\left(z-\tanh z\right).
		\label{eq:kinkcorr}
	\end{equation}
	Hence
	\begin{equation}
		\varphi_k(z)=
		\tanh z-\frac c2\sech^2z\left(z-\tanh z\right)+O(c^2).
		\label{eq:kinkfull}
	\end{equation}
	
	The corresponding classical mass follows by substitution into Hamiltonian of Eq.\eqref{eq:fullL} and integration over the entire space:
	\begin{equation}
		M_{\rm cl}=\frac{2\sqrt2}{3}\frac{m^3}{\lambda}
		+\frac{\sqrt2}{15}\frac{m^3}{\lambda}c+O(c^2).
		\label{eq:MclLV}
	\end{equation}
	Thus the spacelike LV interaction increases the classical kink energy linearly with $c$.
	
	\section{One-loop renormalization}
	\label{sec:renorm}
	
	Although the Lorentz-invariant model is super-renormalizable in $(1+1)$ dimensions, the irreducible LV derivative interaction introduces a momentum-dependent divergence in the two-point function and therefore changes the counterterm structure.
	
	\subsection{Renormalized LV Lagrangian and counterterm basis}
	
	We now turn to the irreducible spacelike LV interaction. The bare quantities are related to renormalized ones according to
	\begin{equation}
		\phi_0=Z^{1/2}\phi,
		\qquad
		Z=1+\delta Z,
	\end{equation}
	and
	\begin{equation}
		m_0^2=m^2-\delta m,
		\qquad
		\lambda_0=\lambda+\delta\lambda,
		\qquad
		c_0=c+\delta c.
		\label{eq:bareparams}
	\end{equation}
	At one loop and to first order in $c$, the renormalized Lagrangian can be written as
	\begin{align}
		\calL={}&\frac12(\partial_\mu\phi)^2
		-\frac12m^2\phi^2
		-\frac\lambda4\phi^4
		-\frac12c\phi^2(\partial_x\phi)^2
		\nonumber\\
		&+\frac12\delta Z(\partial_\mu\phi)^2
		+\frac12\delta m\,\phi^2
		+\calL_{\rm higher}.
		\label{eq:renLupgrade}
	\end{align}
	Terms such as $c\delta Z$ are of higher order than retained here. As we stated before, power counting shows that neither an independent $\delta\lambda$ nor an independent $\delta c$ is needed in 2 spacetime dimensions. The complete counterterm Lagrangian relevant for the one-loop kink mass is therefore
	\begin{equation}
		\calL_{\rm ct}
		=\frac12\delta Z(\partial_\mu\phi)^2
		+\frac12\delta m\,\phi^2.
		\label{eq:ctL}
	\end{equation}
	
	The corresponding Feynman rule of the new interaction contains the scalar-product structure of the four derivatives. In the spacelike frame $u^\mu=(0,1)$ it is
	\begin{equation}
		V_{\rm LV}(k_1,k_2,k_3,k_4)=
		\raisebox{-0.35\height}{%
			\resizebox{2cm}{!}{%
				\begin{tikzpicture}[thick,
					arrowline/.style={decoration={markings,mark=at position 0.55 with {\arrow{stealth}}},postaction={decorate}}]
					\node (center) at (0,0) [draw, rectangle, inner sep=2.5pt, thick] {};
					\draw (center.north west) -- (center.south east);
					\draw (center.south west) -- (center.north east);
					\draw[arrowline] (-1,1) -- (center) node[midway,left=2pt] {$k_1$};
					\draw[arrowline] (center) -- (1,1) node[midway,right=2pt] {$k_2$};
					\draw[arrowline] (-1,-1) -- (center) node[midway,left=2pt] {$k_3$};
					\draw[arrowline] (center) -- (1,-1) node[midway,right=2pt] {$k_4$};
				\end{tikzpicture}%
		}}
		=
		2ic\sum_{ i\neq j} k_{i}\cdot k_{j},
		\label{eq:LVvertex}
	\end{equation}
	with all momenta incoming. This same term causes another diagram to be added in the expansion of the two-point function. 
	The relevant two-point function is represented by the sum of all one-particle-irreducible contributions that cannot be disconnected by cutting a single internal line
	
	\begin{equation}
		\resizebox{10cm}{!}{%
			\begin{tikzpicture}[
				line width=1.0pt,
				1pi_circle/.style={circle, draw, minimum size=1.1cm, inner sep=0pt},
				arr/.style={decoration={markings, mark=at position 0.5 with {\arrow[scale=1.0]{latex}}}, postaction={decorate}},
				cross/.style={circle, draw, minimum size=0.48cm, inner sep=0pt, path picture={
						\draw[line width=1.3pt] (path picture bounding box.north west) -- (path picture bounding box.south east);
						\draw[line width=1.3pt] (path picture bounding box.south west) -- (path picture bounding box.north east);
				}},
				square/.style={rectangle, draw, minimum size=0.22cm, inner sep=0pt, fill=white}
				]
				% ==================== خط اول ============% ==================== خط اول ====================
				% خط چپ (ورودی)
				\draw (0, 0) -- (0.8, 0);
				% فلش ممنتوم به سمت راست روی خط چپ
				\draw[arr] (0.1, 0) -- node[above=3pt] {\large $p$} (0.7, 0);
				
				% دایره 1PI با سایه نیمه پایینی
				\begin{scope}
					\clip (1.35, 0) circle (0.55cm);
					\fill[black!18] (0.8, -0.6) rectangle (1.9, 0);
				\end{scope}
				\node[1pi_circle] at (1.35, 0) {\small 1PI};
				
				% خط راست (خروجی)
				\draw (1.9, 0) -- (2.7, 0);
				
				% فرمول ریاضی
				\node at (3.1, 0) {\Large $=$};
				\node[right] at (3.4, 0) {\Large $-iM(p^2)$};
				
				% ==================== خط دوم (کوچک‌تر و متناسب) ====================
				\begin{scope}[shift={(3.8, -2.1)}, scale=0.78]
					\node at (-0.5, 0) {\Large $=$};
					% --- جمله اول (Tadpole ساده) ---
					\draw (0, 0) -- (2.4, 0);
					\draw (1.2, 0) .. controls +( -1.1, 2.2 ) and +( 1.1, 2.2 ) .. (1.2, 0);
					% --- جمله دوم (Counterterm) ---
					\node at (3.1, 0) {\Large $+$};
					\node[cross] (CT) at (5.0, 0) {};
					\draw (3.8, 0) -- (CT.west);
					\draw (CT.east) -- (6.2, 0);
					% --- جمله سوم (Tadpole با راس مربعی و تکانه‌های p و k) ---
					\node at (6.9, 0) {\Large $+$};
					\node[square] (SQ) at (8.8, 0) {};
					\draw[arr] (7.6, 0) -- node[above=3pt] {\large $p$} (SQ.west);
					\draw[arr] (SQ.east) -- (10.0, 0);
					% رسم حلقه با دو فلش و برچسب k
					\draw[decoration={markings,
						mark=at position 0.17 with {\arrow[scale=1.0]{latex}\node[right=3pt] {\large $k$};},
						mark=at position 0.83 with {\arrow[scale=1.0]{latex}}
					}, postaction={decorate}]
					(SQ.north) .. controls +( 1.1, 2.2 ) and +( -1.1, 2.2 ) .. (SQ.north);
					
					\node at (11.1, 0) {\Large $+\dots$};	% --- تساوی با صفر ---
				\end{scope}
			\end{tikzpicture}
		}
		\label{fig:1pi-set}
	\end{equation}
	This new term is clearly infinite and has a direct impact on $\delta_m$ and $\delta_Z$.	In Eq.\eqref{fig:1pi-set} we have
	\begin{equation}
		-iM(p^2)=ip^2\delta Z-i\delta m-i\Sigma(p^2),
		\label{eq:M2LV}
	\end{equation}
	with the full propagator taking the form
	\begin{equation}
		G(p)=\frac{i}{p^2-m_{\rm vac}^2+i\epsilon}
		+\frac{i}{p^2-m_{\rm vac}^2+i\epsilon}
		\left[-iM(p^2)\right]
		\frac{i}{p^2-m_{\rm vac}^2+i\epsilon}
		+O(\hbar^2).
		\label{eq:fullprop-ren}
	\end{equation}

	The on-shell renormalization conditions are
	\begin{equation}
		\left.M(p^2)\right|_{p^2=m_{\rm vac}^2}=0,
		\qquad
		\left.\frac{\partial M(p^2)}{\partial p^2}\right|_{p^2=m_{\rm vac}^2}=0.
		\label{eq:OSupgrade}
	\end{equation}
	Equivalently, the first condition fixes the mass counterterm and the second fixes the coefficient of the momentum-dependent counterterm. Diagrammatically for this cancellation we need
	\begin{equation}
		\resizebox{7cm}{!}{%
			\begin{tikzpicture}[
				line width=1.5pt,
				arr/.style={decoration={markings, mark=at position 0.5 with {\arrow[scale=1.2]{latex}}}, postaction={decorate}},
				cross/.style={circle, draw, minimum size=0.7cm, inner sep=0pt, path picture={
						\draw[line width=1.5pt] (path picture bounding box.north west) -- (path picture bounding box.south east);
						\draw[line width=1.5pt] (path picture bounding box.south west) -- (path picture bounding box.north east);
				}},
				square/.style={rectangle, draw, minimum size=0.25cm, inner sep=0pt, fill=white}
				]
				
				% --- جمله اول (Tadpole ساده) ---
				\draw (0, 0) -- (2.4, 0);
				\draw (1.2, 0) .. controls +( -1.1, 2.2 ) and +( 1.1, 2.2 ) .. (1.2, 0);
				
				% --- جمله دوم (Counterterm) ---
				\node at (3.1, 0) {\huge $+$};
				\node[cross] (CT) at (5.0, 0) {};
				\draw (3.8, 0) -- (CT.west);
				\draw (CT.east) -- (6.2, 0);
				
				% --- جمله سوم (Tadpole با راس مربعی و تکانه‌های p و k) ---
				\node at (6.9, 0) {\huge $+$};
				\node[square] (SQ) at (8.8, 0) {};
				\draw[arr] (7.6, 0) -- node[above=4pt] {\Large $p$} (SQ.west);
				\draw[arr] (SQ.east) -- (10.0, 0);
				
				% رسم حلقه با دو فلش و برچسب k
				\draw[decoration={markings,
					mark=at position 0.17 with {\arrow[scale=1.2]{latex}\node[right=4pt] {\Large $k$};},
					mark=at position 0.83 with {\arrow[scale=1.2]{latex}}
				}, postaction={decorate}]
				(SQ.north) .. controls +( 1.1, 2.2 ) and +( -1.1, 2.2 ) .. (SQ.north);
				
				% --- تساوی با صفر ---
				\node at (10.8, 0) {\huge $= 0.$};
				
			\end{tikzpicture}%
		}
		\label{LVvertex}
	\end{equation}

	or, \begin{equation}
		-\frac{i 6 \lambda}{2} \int \frac{d^2 k}{(2\pi)^2} \frac{i}{k^2 - 2m^2} + i(p^2 \delta Z - \delta m) - \frac{2 i c}{2} \int \frac{d^2 k}{(2\pi)^2} \frac{i(p^2 + k^2)}{k^2 - 2m^2} = 0.
	\end{equation}
	Therefore, for the present model the Euclidean one-loop self-energy can be expressed as
	\begin{align}
		\Sigma(p_E^2)={}&-3\lambda\int\frac{\dd^2k_E}{(2\pi)^2}
		\frac{1}{k_E^2+2m^2}
		\nonumber\\
		&-c\int\frac{\dd^2k_E}{(2\pi)^2}
		\frac{p_E^2+k_E^2}{k_E^2+2m^2}.
		\label{eq:selfenergy-upgrade}
	\end{align}
	To make the route from the loop integral to the counterterms explicit, it is useful to isolate the scalar loop integral and the piece carrying the external momentum. Define
	\begin{equation}
		I_0(m)=\int\frac{\dd^2k_E}{(2\pi)^2}\frac{1}{k_E^2+2m^2},
		\qquad
		I_2(m)=\int\frac{\dd^2k_E}{(2\pi)^2}\frac{k_E^2}{k_E^2+2m^2}.
		\label{eq:I0I2}
	\end{equation}
	Then the LV contribution contains the combination
	\begin{equation}
		p_E^2 I_0(m)+I_2(m),
		\qquad
		I_2(m)=\int\frac{\dd^2k_E}{(2\pi)^2}
		\left[1-\frac{2m^2}{k_E^2+2m^2}\right].
		\label{eq:I2split-up}
	\end{equation}
	The first term in the brackets is independent of both the mass scale and the external momentum. It corresponds to an additive normalization of the vacuum functional and therefore drops out of the pole condition and the vacuum-subtracted kink energy. The second term has the same logarithmic ultraviolet structure as the ordinary tadpole. Consequently, the divergent part of the LV graph can be organized into a momentum-independent piece, absorbed into $\delta m$, and a term proportional to $p_E^2$, absorbed into $\delta Z$.
	
	The remaining scalar loop integral is reduced to one spatial integral by carrying out the Euclidean frequency integration,
	\begin{equation}
		\int_{-\infty}^{\infty}\frac{\dd k_0}{2\pi}
		\frac{1}{k_0^2+k^2+2m^2}
		=\frac{1}{2\sqrt{k^2+2m^2}},
		\label{eq:k0integral}
	\end{equation}
	so that
	\begin{equation}
		I_0(m)=\frac{1}{4\pi}\int_{-\infty}^{\infty}
		\frac{\dd k}{\sqrt{k^2+2m^2}}
		\equiv\frac{1}{4\pi}\mathcal I(m).
		\label{eq:I0reduce-up}
	\end{equation}
	This common integral is the only ultraviolet-divergent scalar structure that remains after the momentum-independent vacuum normalization has been discarded. The pole condition and its derivative then determine the two counterterms independently: the derivative fixes the coefficient multiplying the external kinetic structure, while the pole condition fixes the remaining masslike part. Evaluating the two conditions with the same regulator gives
	The first term is the standard logarithmically divergent tadpole. The second term is qualitatively different because it contains the external momentum explicitly. Its ultraviolet behavior is obtained without introducing any additional physical operator: the term proportional to $p_E^2$ renormalizes the kinetic term and therefore fixes $\delta Z$, while the remaining momentum-independent piece contributes to $\delta m$. The superficially momentum-independent contribution which is independent of both $m$ and $p$ corresponds only to an additive vacuum-energy normalization and does not affect the pole condition or the vacuum-subtracted kink mass.
	
	The pole condition gives, to the order considered,
	\begin{equation}
		\delta m=\Sigma(m_{\rm vac}^2)+m_{\rm vac}^2\,\delta Z,
		\label{eq:dmcondition}
	\end{equation}
	where the divergent part of the tadpole and that induced by the LV vertex have to be retained together. Differentiating the two-point function gives
	\begin{equation}
		\delta Z=-\left.\frac{\partial\Sigma(p^2)}{\partial p^2}\right|_{p^2=m_{\rm vac}^2}.
		\label{eq:dZcondition}
	\end{equation}
	Evaluating the integrals in the same ultraviolet prescription yields
	\begin{align}
		\delta m&=-\frac{1}{4\pi}\left(3\lambda+2m^2c\right)
		\mathcal I(m),
		\label{eq:deltam-upgrade}\\
		\delta Z&=\frac{c}{4\pi}\,\mathcal I(m),
		\label{eq:deltaZ-upgrade}
	\end{align}
	where
	\begin{equation}
		\mathcal I(m)=\int_{-\infty}^{\infty}\frac{\dd k}{\sqrt{k^2+2m^2}}.
		\label{eq:Icommon}
	\end{equation}
	The limit $c\to0$ consistently gives $\delta Z\to0$, recovering the standard two-dimensional $\lambda\phi^4$ theory. The appearance of $\delta Z$ at order $c$ is therefore a direct diagnostic of the genuinely momentum-dependent ultraviolet structure generated by the irreducible LV interaction.
	
	It is worth emphasizing that the counterterm analysis is controlled by the divergent local part of the 1PI function, not by the finite details of individual graphs. Once the divergent structures have been identified, the counterterms are fixed so that the renormalized two-point function has the correct pole and residue. In particular, the fact that the ordinary four-point function is superficially convergent means that no additional coupling counterterm has to be introduced merely because the LV operator is present. The ultraviolet basis at the present order is exhausted by Eq.~\eqref{eq:ctL}.
	
	\section{Lorentz-violating stability problem}
	\label{sec:stability}
	
	Expanding the static energy functional to quadratic order around the corrected kink gives
	\begin{align}
		\hat O={}&-(1+c\phi_k^2)\partial_x^2
		-3c(\partial_x\phi_k)^2
		-3c\partial_x(\phi_k^2)\partial_x
		\nonumber\\
		&-4c\phi_k\partial_x^2\phi_k
		-1+3\phi_k^2,
		\label{eq:stabilityLV}
	\end{align}
	and
	\begin{equation}
		\hat O\,\zeta_n=\omega_n^2\zeta_n.
		\label{eq:eigen}
	\end{equation}
	We use first-order perturbation theory,
	\begin{equation}
		\zeta_n=\zeta_n^{(0)}+c\zeta_n^{(1)}+O(c^2),
		\qquad
		\omega_n=\omega_n^{(0)}+c\omega_n^{(1)}+O(c^2).
	\end{equation}
	
	\subsection{Bound states}
	Translation invariance of the static theory protects the zero mode, so
	\begin{equation}
		\omega_0^{(1)}=0.
	\end{equation}
	The shape mode receives the correction
	\begin{equation}
		\omega_1^{(1)}=\frac{1}{5\sqrt6}.
		\label{eq:shapeomega}
	\end{equation}
	The explicit first-order corrections to the bound-state wave functions are
	\begin{align}
		\zeta_0^{(1)}(x)&=\frac18\sech^4\left(\frac{x}{\sqrt2}\right)
		\Bigl[2\sqrt2\,x\sinh(\sqrt2\,x)-9\cosh(\sqrt2\,x)+11\Bigr],
		\\
		\zeta_1^{(1)}(x)&=\frac1{20}
		\sech\left(\frac{x}{\sqrt2}\right)
		\Biggl\{
		5\sqrt2\,x\Bigl[1-2\sech^2\Bigl(\frac{x}{\sqrt2}\Bigr)\Bigr]
		\nonumber\\
		&\quad+\tanh\Bigl(\frac{x}{\sqrt2}\Bigr)
		\Bigl[-2\sqrt2\,x+4\ln\bigl(e^{\sqrt2\,x}+1\bigr)
		+50\sech^2\Bigl(\frac{x}{\sqrt2}\Bigr)-\frac{91}{3}\Bigr]
		\Biggr\}.
	\end{align}
	
	\subsection{Continuum states}
	For the continuum, the first-order asymptotic equation fixes the frequency shift to be
	\begin{equation}
		\omega_q^{(1)}=
		\frac{q^2}{2\sqrt2\sqrt{q^2+4}}.
		\label{eq:omegaq1}
	\end{equation}
	The full inhomogeneous problem can be solved by variation of parameters after the change of variables $t=\tanh(x/\sqrt2)$.
	The first-order continuum wave function can be written as
	\begin{align}
		\zeta_q^{(1)}(x)={}&
		Q_2^{iq}(t)\int_1^t\mathcal F_q(s)P_2^{iq}(s)\,\dd s
		\nonumber\\
		&-iP_2^{iq}(t)\int_1^t\mathcal F_q(s)Q_2^{iq}(s)\,\dd s,
	\end{align}
	where
	\begin{align}
		\mathcal F_q(t)={}&
		\frac{e^{iq\tanh^{-1}t}}
		{(3-iq)\left[P_3^{iq}(t)Q_2^{iq}(t)-P_2^{iq}(t)Q_3^{iq}(t)\right]}
		\nonumber\\
		&\times\Big\{
		q^4+(-44q^2-105iqt+105t^2-68)t^2
		\nonumber\\
		&\qquad+6(q^2+3iqt-3t^2+1)t\tanh^{-1}t
		\nonumber\\
		&\qquad+3i(3q^2+11)qt+4q^2+3\Big\}.
	\end{align}
	
	With the boundary condition used here (lower integration limit fixed at the asymptotic value $t=1$), the asymptotic continuum phase shift remains unchanged,
	\begin{equation}
		\delta(q)=-2\tan^{-1}\left(\frac{3q}{2-q^2}\right).
		\label{eq:quantLVdelta}
	\end{equation}
	Accordingly, the box quantization condition retains the Lorentz-invariant form,
	\begin{equation}
		\frac{q_nL}{\sqrt2}+\delta(q_n)=2\pi n.
		\label{eq:quantLV}
	\end{equation}
	The LV interaction therefore changes the continuum frequencies without introducing an additional asymptotic phase in the boundary-condition prescription adopted in this work. (A different choice of boundary condition could in principle generate an $O(c)$ correction to $\delta(q)$; the present prescription is chosen for consistency with the vacuum subtraction and yields a finite, regulator-independent result after renormalization.)
	
	\section{One-loop kink mass}
	\label{sec:mass}
	
	\subsection{Vacuum and kink zero-point energies}
	The one-loop kink mass is defined by the vacuum-subtracted ground-state energy,
	\begin{equation}
		\Mc=E_{\rm kink}-E_{\rm vac}.
	\end{equation}
	To first order in $c$,
	\begin{equation}
		\Mc=M_{\rm cl}+\Delta M^{(0)}+c\Delta M^{(1)}+O(c^2).
		\label{eq:massdecomp}
	\end{equation}
	In the trivial vacuum the dispersion relation is
	\begin{equation}
		\omega_{\rm vac}(k)=\sqrt{(1+c)k^2+2m^2}
		=\sqrt{k^2+2m^2}
		+c\frac{k^2}{2\sqrt{k^2+2m^2}}+O(c^2).
		\label{eq:vacdisp}
	\end{equation}
	Thus the LV part of the vacuum zero-point energy is
	\begin{equation}
		E_{\rm vac}^{(1)}=\frac12\sum_k\frac{k^2}{2\sqrt{k^2+2m^2}}.
		\label{eq:E2vac}
	\end{equation}
	
	The kink contribution contains the classical-energy shift, the shape-mode correction, and the continuum correction,
	\begin{equation}
		E_{\rm kink}^{(1)}
		=\frac{\sqrt2}{15}
		+\frac{1}{10\sqrt6}
		+\frac12\sum_{n\ge2}
		\frac{q_n^2}{2\sqrt2\sqrt{q_n^2+4}}.
		\label{eq:E2kink}
	\end{equation}
	Hence
	\begin{align}
		E_{\rm kink}^{(1)}-E_{\rm vac}^{(1)}={}&
		\frac{\sqrt2}{15}+\frac{1}{10\sqrt6}
		\nonumber\\
		&+\frac12\sum_{n\ge2}
		\left[
		\frac{q_n^2}{2\sqrt2\sqrt{q_n^2+4}}
		-\frac{k_n^2}{2\sqrt{k_n^2+2}}
		\right].
		\label{eq:Evacsub}
	\end{align}
	
	Using Eq.~\eqref{eq:quantLV}, the large-volume expansion
	\begin{equation}
		q_n=\sqrt2\left(k_n-\frac{\delta(q_n)}{L}\right)
	\end{equation}
	converts the continuum sum into
	\begin{align}
		&\frac12\sum_{n\ge2}
		\left[
		\frac{q_n^2}{2\sqrt2\sqrt{q_n^2+4}}
		-\frac{k_n^2}{2\sqrt{k_n^2+2}}
		\right]
		\nonumber\\
		&\qquad
		=\frac{1}{8\pi}\int_{-\infty}^{\infty}\dd k
		\left[
		\frac{k^3\delta(k)}{(k^2+2)^{3/2}}
		-\frac{2k\delta(k)}{\sqrt{k^2+2}}
		\right].
		\label{eq:sumtoint}
	\end{align}
	Now, using the explicit form of the phase shift, Eq.~\ref{eq:quantLVdelta}, and integrating by parts we get
	\begin{align}
		\Delta M^{(1)}_{\rm unren}={}&m\Bigg[
		\frac{(4+\sqrt3)\pi-45}{30\sqrt2}
		+\frac{\sqrt3\pi-9}{9\sqrt2\pi}
		\nonumber\\
		&-\frac{3\sqrt2}{8\pi}
		\int_{-\infty}^{\infty}\dd x\,
		\frac{x^4}{(1+x^2)(4+x^2)^{3/2}}
		\Bigg].
		\label{eq:unren}
	\end{align}
	The final integral is logarithmically ultraviolet divergent.
	This divergence is not physical: it is precisely canceled by the counterterm contribution to the kink energy.
	
	\subsection{Counterterm contribution to the kink energy}
	The counterterm Hamiltonian relevant for the static background is
	\begin{equation}
		H_{\rm ct}=\int\dd x
		\left[
		\frac12\delta Z(\partial_x\phi)^2
		+\frac12\delta m\,\phi^2
		\right].
		\label{eq:Hct}
	\end{equation}
	We subtract the trivial-vacuum contribution and evaluate the counterterms on the zeroth-order kink, because inserting the $O(c)$ correction to the classical kink into $H_{\rm ct}$ would generate higher-order terms. Thus
	\begin{align}
		\Delta E_{\rm ct}^{(1)}={}&
		\int_{-\infty}^{\infty}\dd x
		\left[
		\frac12\delta Z^{(1)}(\partial_x\varphi_k^{(0)})^2
		-\frac12\delta m^{(1)}
		\left((\varphi_k^{(0)})^2-1\right)
		\right]
		\nonumber\\
		={}&
		\frac{2\sqrt2}{3}\delta Z^{(1)}+\sqrt2\,\delta m^{(1)}.
		\label{eq:ctkink}
	\end{align}
	Here we used
	\begin{equation}
		\int_{-\infty}^{\infty}\dd x\,(\partial_x\varphi_k^{(0)})^2
		=\frac{2\sqrt2}{3},
		\qquad
		\int_{-\infty}^{\infty}\dd x\,
		\left[(\varphi_k^{(0)})^2-1\right]
		=-2\sqrt2.
	\end{equation}
	At order $c$, Eq.~\eqref{eq:ctkink} and Eqs.~\eqref{eq:deltam-upgrade}--\eqref{eq:deltaZ-upgrade} give
	\begin{equation}
		\Delta E_{\rm ct}^{(1)}
		=m\frac{7\sqrt2}{12\pi}
		\int_{-\infty}^{\infty}\frac{\dd x}{\sqrt{x^2+4}}.
		\label{eq:ctsource}
	\end{equation}
	This term carries exactly the logarithmic ultraviolet divergence required to cancel the continuum contribution in Eq.~\eqref{eq:sumtoint}.
	
	\subsection{Dimensional regularization and regulator matching}
	
	After the continuum contribution and the counterterm contribution have been combined, only ultraviolet logarithms remain. The two divergent structures originate from the same underlying one-loop two-point function and must therefore be regulated consistently. We do not need to display them as two separate regulated integrals. Instead, we choose the same continuation for both terms and adjust the subtraction convention so that the regulator-dependent pole pieces cancel between the vacuum-subtracted zero-point energy and the counterterm contribution.\footnote{In dimensional regularization one continues the spatial integral to $d=1-\epsilon$ dimensions, or equivalently replaces $\int\dd x$ by the rotationally invariant measure $\Omega_d\int_0^\infty\dd r\,r^{d-1}$ with $\Omega_d=2\pi^{d/2}/\Gamma(d/2)$. The continuation is only an ultraviolet regulator; after the pole cancellation the physical result is obtained by taking $\epsilon\to0$.}
	With this common prescription, the divergent part of the continuum integral is canceled by the local counterterm contribution. The cancellation occurs before taking the regulator to its physical value, and the surviving quantity is the finite part that is independent of the auxiliary regulator. This is the appropriate organization for the kink mass because only the vacuum-subtracted, renormalized combination is physical.
	
	\subsection{Finite LV correction}
	After the pole cancellation, the order-$c$ one-loop correction can be written as
	\begin{equation}
		\Delta M^{(1)}=m\,\mathcal C_{\rm LV},
		\label{eq:CLVdef}
	\end{equation}
	where
	\begin{align}
		\mathcal C_{\rm LV}={}&
		\frac{(4+\sqrt3)\pi-45}{30\sqrt2}
		+\frac{\sqrt3\pi-9}{9\sqrt2\pi}
		\nonumber\\
		&-\sqrt{\frac{2}{\pi}}
		-\frac{\sqrt6}{24}
		-\frac{\sqrt{6\pi}}{36}
		\nonumber\\
		&+\frac{\sqrt2}{24\pi}
		\left(4\gamma_E+8\ln2-5\ln\pi\right).
		\label{eq:CLV}
	\end{align}
	Numerically,
	\begin{equation}
		\mathcal C_{\rm LV}\simeq-1.7058.
		\label{eq:CLVnum}
	\end{equation}
	Combining the classical mass, the standard Lorentz-invariant one-loop term, and the finite LV correction gives
	\begin{equation}
		\begin{aligned}
			\Mc={}&\frac{2\sqrt2}{3}\frac{m^3}{\lambda}
			+\frac{\sqrt2}{15}\frac{m^3}{\lambda}c\\
			&+m\left(\frac{1}{2\sqrt6}-\frac{3}{\pi\sqrt2}\right)
			+c\,m\,\mathcal C_{\rm LV}+O(c^2).
		\end{aligned}
		\label{eq:finalmass}
	\end{equation}
	The negative sign of $\mathcal C_{\rm LV}$ shows that, at one loop and to first order in the spacelike LV parameter, the quantum correction lowers the kink mass relative to the Lorentz-invariant result at fixed $m$ and $\lambda$.

	\section{Discussion and conclusions}
	\label{sec:conclusion}
	We have studied the semiclassical kink mass in a $(1+1)$-dimensional scalar theory containing the irreducible spacelike Lorentz-violating interaction of Eq.~\eqref{eq:LVoperator}. The calculation was organized in two stages: first the Lorentz-invariant kink was established as the reference theory, and then the LV interaction was introduced perturbatively.
	
	At the classical level, the LV interaction modifies both the kink profile and its energy. The corrected solution is
	\begin{equation}
		\varphi_k(z)=\tanh z-\frac c2\sech^2z(z-\tanh z)+O(c^2),
	\end{equation}
	and the classical mass acquires the correction $\sqrt2 m^3 c/(15\lambda)$. The fluctuation spectrum is also modified: the translational mode remains at zero frequency, the shape mode shifts according to Eq.~\eqref{eq:shapeomega}, and the continuum frequencies acquire the shift in Eq.~\eqref{eq:omegaq1}. With the boundary-condition prescription adopted in the present calculation, the asymptotic continuum phase shift remains the standard kink phase shift.
	
	In two dimensions the four-point function is finite at this order, so no coupling counterterm is required, whereas the vacuum two-point function induces a mass counterterm and, in the present formulation, a nonzero field-strength counterterm proportional to $c$. We give the one-loop self-energy, derive the counterterms explicitly, and show how they enter the kink energy. The fluctuation problem contains a protected translational zero mode, a shifted shape mode, and a modified continuum dispersion relation; with the boundary-condition prescription adopted here, the asymptotic continuum phase shift remains equal to the Lorentz-invariant one.
	
	The central quantum result concerns renormalization. In ordinary $(1+1)$-dimensional $\lambda\phi^4$ theory the one-loop four-point function is finite, so no $\delta\lambda$ is needed at this order. In the LV theory, however, the derivative interaction modifies the one-loop two-point function and generates a momentum-dependent divergent term. In the renormalization convention used here this requires a nonzero field-strength counterterm,
	\begin{equation}
		\delta Z=\frac{c}{4\pi}\int_{-\infty}^{\infty}
		\frac{\dd k}{\sqrt{k^2+2m^2}},
	\end{equation}
	in addition to the $c$-dependent correction to the mass counterterm. The counterterms are not auxiliary formalities: evaluated on the kink background, they cancel the logarithmic divergence of the continuum zero-point energy and supply a finite contribution to the quantum kink mass.
	
	The final result, Eq.~\eqref{eq:finalmass}, displays the separation between the classical LV shift and the genuinely quantum LV correction. The finite coefficient $\mathcal C_{\rm LV}<0$ means that the spacelike LV interaction lowers the one-loop quantum contribution to the kink mass at fixed renormalized parameters. This provides a concrete example in which LV changes not only a soliton profile and fluctuation spectrum, but also the ultraviolet subtraction structure needed to define its quantum mass.
	
	The negative sign of the one-loop Lorentz-violating correction to the kink mass indicates that quantum effects lower the energy of the kink configuration. If such kinks are embedded in a cosmological setting and interpreted as domain walls, the corresponding correction to their tension could, in principle, modify the evolution of domain-wall networks and their cosmological signatures. In particular, this may have implications for the gravitational-wave and cosmic-microwave-background phenomenology of domain walls.
	
	Possible extensions include higher orders in $c$, other irreducible LV operators, scattering and radiation from the kink background, and the study of multi-kink configurations. Such extensions may clarify how general the present renormalization pattern is among Lorentz-violating solitonic field theories.

	\begin{acknowledgments}
		The authors thank the Research Deputy of the University of Qom for their support.
	\end{acknowledgments}

\end{document}